\documentclass[aps,preprint,superscriptaddress]{revtex4-1}

\usepackage{amsmath,amssymb,amsfonts}
\usepackage{epsfig,graphicx,color}

\newcommand{\bfe}{{\boldsymbol e}}%
\newcommand{\bfm}{{\boldsymbol m}}%
\newcommand{\bfs}{{\boldsymbol s}}%
\newcommand{\bft}{{\boldsymbol t}}%
\newcommand{\bfW}{{\boldsymbol W}}%
\newcommand{\bfx}{{\boldsymbol x}}%
\newcommand{\bfX}{{\boldsymbol X}}%
\newcommand{\bfy}{{\boldsymbol y}}%

\newcommand{\bfsigma}{{\boldsymbol \sigma}}%

\newcommand{\rr}{{\mathbb R}}%
\newcommand{\R}{{\mathbb R}} %
\newcommand{\mcC}{{\mathcal{C}}}%
\newcommand{\mcI}{{\mathcal{I}}}%

\newcommand{\bfzero}{{\boldsymbol 0}}%
\newcommand{\bftau}{{\boldsymbol \tau}}%
\newcommand{\bfphi}{{\boldsymbol \varphi}}%
\newcommand{\bbP}{{\mathbb{P}}}%

\DeclareMathOperator*{\argmin}{arg\,min} % thin space, limits underneath in displays

\begin{document}
\title{Minimum energy paths  for conformational changes of viral capsids} 

\author{Paolo Cermelli}
 \email{paolo.cermelli@unito.it}
\affiliation{Dipartimento di Matematica, Universit\`a di Torino, Italy}

\author{Giuliana Indelicato}
\email{giuliana.indelicato@unito.it}
\affiliation{Department of Mathematics and York Centre for Complex Systems Analysis, University of York, UK and Dipartimento di Matematica, Universit\`a di Torino, Italy }

\author{Emilio Zappa}
\email{zappa@cims.nyu.edu}
\affiliation{Courant Institute of Mathematical Sciences, New York University, USA}

\begin{abstract} In this work we study conformational changes of viral capsids using techniques of Large Deviations Theory for stochastic differential equations. The viral capsid is a model of a complex system in which many units - the proteins forming the capsomers -  interact by weak forces to form a structure with exceptional mechanical resistance. The destabilization of such a structure is interesting both {\it per se}, since it is related either to infection or maturation processes, and because it yields insights into the stability of complex structures in which the constitutive elements interact by weak attractive forces.  We focus here on a simplified model of a dodecahedral viral capsid, and assume that the capsomers are rigid plaquettes with one degree of freedom each. We compute the most probable transition path from the closed capsid to the final configuration using minimum energy paths, and discuss the stability of intermediate states. 
\end{abstract}


\maketitle

\section{Introduction}

Viral capsids are interesting biological structures assembled from repeated copies of the same protein \cite{crick}. They are very efficient at their purpose of protecting the genetic material of the virus  from the environment, since they are quite stable for a wide range of environmental conditions (cf. \cite{zandi_2} for a review).

In order to release the genome inside the host cell, however,  the capsid must be able to change configuration and /or disassemble in response to changes in the chemical environment or the interaction with receptors of the host. Such conformational changes often involve the opening of pores in the viral shell through which the nucleic acid exits the virus and is released into the cell (see, for instance, \cite{tuthill, bakker,blaas2000,blaas2002,blaas2012}).  More often, though, the capsid is believed to simply disassemble as a consequence of the weakening of the 
bonds between the capsomers \cite{cao}.

Also, many viral capsids undergo structural changes during maturation \cite{hk97_2}. The assembly of the capsid is often a multi-stage process that may involve various steps towards the infective, final form of the virion. Once the capsomers have assembled to form a closed shell, called  procapsid, these  still have to undergo conformational changes involving protein cleavage, subunit rotation and /or deformation, and substantial bond disruption and reforming, in order to reach the final, stable form of the infective virus \cite{conway}.

Hence, the definition of conformational change is necessarily somewhat loose, ranging from the simple mutual detachment of the proteins leading to the complete disgregation of the capsid, to the cleavage of the capsomers triggering  complex relative rigid motions of the protein subunits, to the formation of new bonds with changes in the quaternary structure of the proteins.

In any event, the question arises as to which is the basic physics underlying the stability and structural plasticity of the capsid. The forces driving the conformational changes are diverse, and not always known. For instance, the nucleic acid is highly compressed within the closed capsid, and this generates an internal pressure that could destabilize the shell \cite{conway},  even though in RNA viruses its large negative charge might contribute to the stability of the capsid through Coulomb attraction with the coat proteins. In other cases, there are charges on the faces of adjacent proteins that are masked by ions in the stable, closed capsid but, upon pH changes, lose the ions and trigger a Coulomb repulsion between the proteins \cite{speir}.  Sometimes, instead, it is the cleavage of a particular protein that acts as switch for a conformational change of the capsomers that reach a new stable configuration \cite{castellanos}.

Cohesive forces that keep the capsid together are also diverse: there are no covalent bonds between the capsomers, and binding usually involves strong hydrophobic bonds between chains 'knotted' around or inside beta barrels of adjacent proteins,  hydrogen bonds, Coulomb attraction, and so on \cite{roos}. 

What is clear, though, is the cooperative nature of the stability of the capsid. The bonds keeping the capsomer together must be comparatively weak in order to allow  for easy and fast configurational changes, and the secret of the capsid stability must lie in the joint stabilizing action of adjacent capsomers upon each other \cite{bakker, hk97_2, castellanos, hk97_1, Zlotnick2003,Zlotnick2009}.

In this work we continue the investigation started in \cite{CIT,giuliana1} (cf. also (\cite{bruinsma2015})), to study under which conditions configurational changes in viral capsids involve either simultaneous collective movements, or completely disordered unrelated events, or a cascade of local destabilization events leading to a wavefront propagating along the shell. Indeed, the energy cascade hypothesis has been  proposed in \cite{ross} and experimentally demonstrated in \cite{cardone} for the maturation of HK97, a complex process that involves multiple (possibly icosahedral) intermediates.

We employ a very simple model of a dodecahedral viral capsid, in which the capsomers are pentagonal rigid units endowed with a single degree of freedom, that may represent a geometric variable, such as a displacement or a rotation of the capsomer as in ERAV and CCMV, or an internal variable measuring the conformation of the protein subunits as in HRV or HK97.  Due to its simplicity, the dodecahedral capsid model is widely used in the literature to perform simulations, but its use here is motivated by the ERAV capsid \cite{bakker}, a pseudo $T=3$ capsid in which disassembly does occur by the relative motion of almost rigid pentagonal units made of 20 coat proteins. Our approach could also be applied to larger, more complex  capsids, in which both the structure and the protein-protein interactions are richer. The purpose here is to put the energy cascade hypothesis on a firm theoretical ground in the simplest  setting as possible.

We assume that   the configurational change is the result of the competition between a  driving force,  that that we view either as the internal pressure due to the confinement of the genetic material inside the capsid, or the Coulomb repulsion between charges at the capsomer-capsomer interfaces, and a counteracting interaction term describing the attraction between adjacent capsomers, consistent with the form of capsid energy proposed in \cite{kegel2004} for HBV.  Notwithstanding the extreme simplicity of  the interaction forces,   the geometry of the interactions between adjacent capsomers  is responsible for the high complexity of the energy landscape, that has many minima corresponding to metastable states of the capsid.

As mentioned above, conformational changes can be triggered by a number of diverse factors. In this work we focus on transitions driven by changes in the chemical enviroment of the capsid that modulate the relative intensity of the two main energetic contributions to capsid stability: for instance, the removal of ions due to changes of the pH usually unmasks charges that are responsible for augmented Coulomb repulsion between the interfaces. All these factors are reckoned by a single parameter $\gamma$ in the energy that  weighs the relative strength of the competing forces, and is such that when it is above a certain threshold, the original configuration of the capsid is stable (immature provirion in maturation events or closed capsid in the disgregation case), and when it decreases below this threshold the initial configuration is unstable. We show that when the critical threshold is reached, the transition between the initial and the final state occurs directly, without intermediates, and requires a large energy \cite{CIT}. Hence, we assume that the parameter $\gamma$ is near but above the critical threshold, in order to retain the fine features of the inter-capsomer interactions in our model.  

In order to explore the energy landscape and determine  the minimum energy paths between metastable states, we use here the formalism of Large Deviations Theory \cite{freidlin}.  The formalism provides both a solid theoretical framework to study transitions, and a numerical procedure to characterize the most probable paths \cite{VDE1, VDE2, VDE3, kohn}.  The idea is to allow for fluctuactions that drive the system out of equilibrium near the transition threshold. The  energy barriers between the minima are related to the times spent by the system in each basin of attraction, and to the probability of transition.

We prove that, under general hypotheses on the energy function, the conformational change occurs by a domino effect in which local destabilization events trigger their neighbors, and this propagates along the capsid until completion.  This confirms the energy cascade hypothesis \cite{ross,cardone} and suggests that local interaction rules govern the details of the transition, which is henceforth not a collective, concerted motion of all capsomers, but rather propagates as a wave along the capsid.

The fact that the transition is an intrinsically local affair is also confirmed by linearizing the model and computing the stationary distribution of fluctuactions around metastable states: the concentration matrix shows that only adjacent pentamers are correlated. Explicit results for a special form of the capsid energy are presented and discussed.

\section{The dodecahedral model}

According to the Caspar and Klug theory,  actual viral capsids are icosahedral shells made of $T$-multiples of 60 proteins, with $T$ an integer \cite{CK}.  The proteins aggregate in small structural units called capsomers, that are usually composed by 2, 3, 5 or 6 proteins. The capsomers are almost rigid, and are often the basic subunits of assembly, as well as of disassembly. Motivated by \cite{bakker}, we focus here on a class of viruses, such as Equine Rhynitis Virus A (ERAV), that release the genome by opening large pores in the capsid by translation and rotation of pentagonal rigid units made of 20 coat proteins. In this case, the capsid behaves as a dodecahedron made of 12 rigid pentagonal faces. The dodecahedon is also a popular model of viral capsid in theoretical investigations, since it is the smallest polyhedron exhibing icosahedral symmetry, still retaining a rich connectivity (each unit is coordinated to 5 other units).

\begin{figure}[h]
\centering
\includegraphics[scale=0.2]{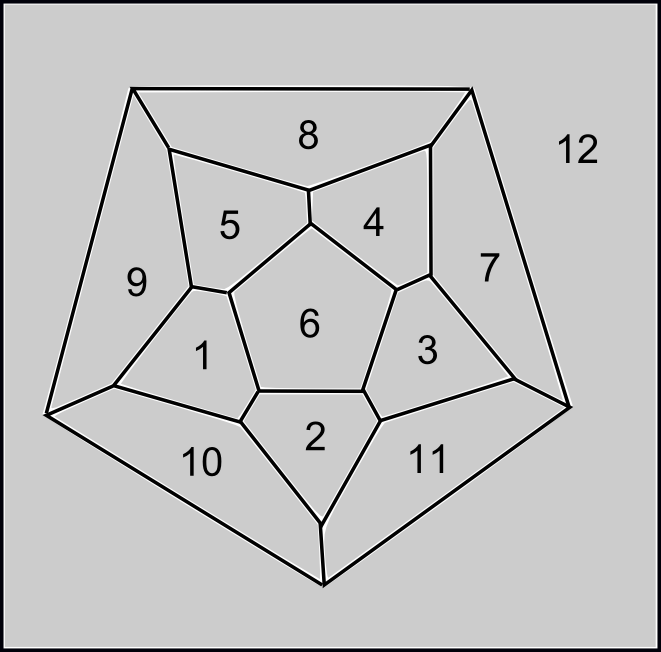}
\caption{Schlegel diagram of the dodecahedron with the indexing convention used here.}
\label{schlegel}
\end{figure}

Hence, in this paper we shall employ a dodecahedron $\mathcal{C}$ (see Figure \ref{schlegel}) as a model of viral capsid. 
Consistently with this simplification, we assume that  every configuration of the capsid is described by an  order parameter $\bfx=(x_1,\dots,x_{12})$,   with indexing corresponding to a labeling of the faces as in Figure \ref{schlegel},  where $x_i\in \rr$ is a variable describing the state of each pentagonal unit. For instance, in destabilization problems  leading to the opening of the capsid during the infection process,  we can assume that each pentagonal face of the dodecahedron $\mathcal{C}$ is a rigid plaquette  that can only translate along an axis orthogonal to its plane, and choose $x_i$ as its radial displacement. 

In order to account for interactions between adjacent pentagons, it is convenient to
work on  the dual graph $G$ of $\mathcal{C}$,  which has the property  that the vertices of $G$ correspond to the faces of $\mathcal{C}$,  the edges of $G$  to the edges of $\mathcal{C}$ and the faces of $G$ to the vertices of $\mathcal{C}$:   $G$ is the graph of the icosahedron.  From now on we view  $\bfx$ as a field on the vertices of the icosahedral graph $G$.

Denoting by $V=\{1,\dots,12\}$ and $E$ the  sets of vertices and edges of $G$, respectively, the adjacency matrix is the square symmetric $12\times 12$ matrix defined by 
 $$
A_{ij}=\left\{
\begin{array}{ll}
1&\text{if  } ij\in E
\\
0&\text{otherwise},
\end{array}
\right.
\qquad
i,j\in\{1,\dots,12\}.
$$
Notice that the $ij$-entry of the adjacency matrix $A$ of $G$ is not vanishing if and only if the $i$ and $j$ faces of the dodecahedron $\mathcal{C}$ meet at a common edge.

\subsection{Symmetry}

We review here some basic relation between the adjacency matrix of $G$ and the symmetry of the associated polyhedron.

We say that a map $V\to V$ is an automorphism of $G$ if it is one-to-one and if it and its inverse maps adjacent vertices into adjacent vertices.  The group of automorphisms of the icosahedral graph $G$ is the Coxeter group $\mathcal{H}_3 = \mcI \times \mathbb{Z}_2$ of order 120, where $\mcI$ denotes the rotational group of the icosahedron, with order 60 \cite{alg_graph}.  The group $\mathcal{I}$ acts on the vertices of $G$, which are the faces of $\mathcal{C}$, inducing a permutation representation (perm rep) $\sigma : \mcI \rightarrow S_{12}$, where $S_{12}$ is the symmetric group over 12 elements. The perm rep $\sigma$ induces a representation $\rho : \mcI \rightarrow GL(12,\rr)$, given by 
\begin{equation*}
\rho(g)  \bfe_j := \bfe_{\sigma(g)(j)},\quad g\in  \mcI,
\end{equation*}
where $\bfe_j, j= 1,\ldots, 12$ denotes the standard basis of $\rr^{12}$. 

 The direct product decomposition of $\mathcal{H}_3$ implies that the representation 
\begin{equation}\label{H3_rep}
\widetilde{\rho} = \rho \otimes \Gamma,
\end{equation}
is a representation of $\mathcal{H}_3$. Here $\Gamma = \{\pm 1 \}$ is a representation of $\mathbb{Z}_2$ and $\otimes$ denotes the tensor product of representations (in this case, since the groups are finite, this is the Kronecker product of matrices) \cite{jones}. Since $\mathcal{H}_3$ is the automorphism group of $G$, $A$ commutes with all the matrices of $\widetilde\rho(\mathcal{H}_3)$.

There is a connection between the eigenspaces of $A$ and the decomposition into irreducible representations (irreps) of $\widetilde{\rho}$ \cite{spectra}. The character table of $\mcI$ is given by 
\begin{center}
\begin{tabular}{l|c c c c c}
Irrep & $C(e)$ & $C(g_{5})$ & $C(g_{5}^{2})$ & $C(g_{2})$ & $C(g_2g_5)$ \\
\hline
$\rho_1$ & 1 & 1 & 1 & 1 & 1 \\
$\rho_2$ & 3 & $\tau$ & 1-$\tau$ & -1 & 0 \\
$\rho_3$ & 3 & 1-$\tau$ & $\tau$ & -1 & 0 \\
$\rho_4$ & 4 & -1 & -1 & 0 & 1 \\
$\rho_5$ & 5 & 0 & 0 & 1 & -1 \\
\end{tabular}
\end{center}
where $\tau = \frac{1+\sqrt{5}}{2}$ is the golden ratio, $g_2$ and $g_5$ a two- and five-fold rotation of the icosahedron, respectively, and $C$ denotes the conjugacy class of an element of the group . The decomposition of $\widetilde{\rho}$ into irreps is given by \cite{ziac} 
\begin{equation*}
\widehat{\rho} = \bigoplus_{i = 1,2,3,5} \rho_i \otimes \Gamma,
\end{equation*} 
and there exists a matrix $R \in GL(12,\rr)$ such that $R^{-1} \widetilde{\rho} R = \widehat{\rho}$. The explicit form of $R$ is given in \cite{ziac}. It is shown in \cite{spectra} that the matrix $R$ diagonalises the adjacency matrix $A$ of the graph $G$; in particular, the spectrum of $A$ is given by
\begin{center}
\begin{tabular}{|c|c|c|}
\hline
Eigenvalue & Dimension & Irrep \\
\hline
$5$ & 1 & $\rho_1$ \\
$\sqrt{5}$ & 3 & $\rho_2$ \\
$-\sqrt{5}$ & 3 & $\rho_3$ \\
$-1$ & 5 & $\rho_5$ \\ 
\hline
\end{tabular}
\end{center}

\section{Energy}

As discussed in the Introduction, we associate with the capsid an energy function $E : \R^{12} \rightarrow \R$ that is the sum of two contributions: a term driving the configurational change, that we identify with the internal pressure due to the confinement of the genomic material inside the capsid,
%or with the unscreened Coulomb repulsion between charges at the pentamer-pentamer interfaces, 
and a term opposing the conformational change, that we identify here with the cohesive force between the pentamers at their interfaces. This assumption is consistent with the proposals of \cite{kegel2004,zandi_2,zandi2006} where it has been used to model assembly.
Hence, we write the total energy of the capsid as
\begin{equation}
E(\bfx) = \sum_{i=1}^{12} f(x_i) + \gamma\sum_{i,j=1}^{12} A_{ij} g(x_i, x_j), \quad \gamma \in  \rr,
\label{energy0}
\end{equation}
with $f:\rr\to\rr$ and $g:\rr^2\to \rr$ to be specified below. The constant $\gamma$ is a real parameter controlling the stability of the capsid, and we assume that it depends on the environment of the capsid, such as pH, salt concentration, temperature.

Notice that, by construction,
the energy is invariant under the action of the symmetry group of the capsid \cite{ziac}. Writing, with a slight abuse of notation, $\widetilde{\rho}(\mathcal{H}_3)\subset GL(12,\rr)$ simply as $\mathcal{H}_3$,
this means that
\begin{equation}\label{energy_inv}
E(H\bfx) = E(\bfx), \qquad \forall H \in  \mathcal{H}_3,\forall \bfx \in  \rr^{12},
\end{equation}
which implies that the gradient $\nabla E(\bfx)$ is equivariant, i.e.
\begin{equation}\label{equivariant}
\nabla (E(H\bfx )) =H \nabla  E(\bfx), \qquad \forall H \in  \mathcal{H}_3.
\end{equation}

We require that $f:\rr\to\rr$ and $g:\rr^2\to \rr$ are smooth and
\begin{itemize}
\item[$A_1$] the expansion energy is monotone decreasing  and convex, and has an absolute minimum at state $x=1$, which we identify with the 'switched' state of the pentamer (either fully detached of switched to the final conformation). Also, we assume that $f(1)=f'(1^-)=0$, $f''(1^-)>0$. For $x>1$ the behavior of $f$ in this model is irrelevant: we can either assume that $f$ is constant, or that it is convex at $f=1$, which implies that the pentamer cannot detach further, or further modify its conformation.  The states for which $x=1$ can be identified to the locked states of \cite{ross}, i.e., intermediate states from which the transition cannot be reversed.

Notice that the convexity of $f$ as a function of a single variable  does not imply the convexity of the energy $E$ as a function of the full state variable $\bfx$. 
%
%Notice that  with the choice of Kegel and Van Der Schoot (formula 3,  \cite{kegel2004}) for the mean potential of repulsion between charges at the protein surfaces. 

\item[$A_2$] the interaction energy $g=g(x,y)$ has a unique strict global mimimum at $x=y=0$, i.e., when both adjacent pentamers are attached to each other or in the initial state, and is symmetric, i.e.,  $g(x,y)=g(y,x)$.   On the other hand, we  must also require that the interaction becomes negligible when two pentamers are either sufficiently far, or one of them is in the switched state. The interaction radius is related to the intermolecular hydrophobic forces between capsomers, and should be much smaller than the distance at which Coulomb interaction is relevant \cite{kegel2004}. 

Hence,   writing 
$g_1=\frac{\partial g}{\partial x}$, $g_2=\frac{\partial g}{\partial y}$, $g_{11}=\frac{\partial^2 g}{\partial x^2}$,  and $g_{12}=\frac{\partial^2 g}{\partial x\partial y}$, we require that $g_1(x,x)\ge0$ and  $g_{12}(x,x)\le 0$ for every $x>0$, and  there is a cutoff value 
$\bar d<<1$ such that $g$ vanishes identically for $x^2+y^2\ge\bar d^2$. In other words, $g(x,y)$ and all its derivatives vanish identically when one of the arguments equals 1. The hypothesis that $g_{12}(x,x)\le 0$ implies that the transition pathway corresponding to $\gamma$ traversing the critical value (see next Section) is isotropic, consistent with normal mode analysis of capsid destabilization \cite{tama}. It is possible to show that this requirement it is satisfied for the potential for hydrophobic interactions in \cite{kegel2004}.
\end{itemize}

We remark that both energies vanish identically when a component $x_i$ is equal to 1. As mentioned above, we identify  states such that $x_i=1$ as states where the corresponding pentamer has switched to its final state, either by changing its conformation, or by detaching from the capsid, depending on the problem at hand. On the other hand, pentamers such that $x_i\sim 0$ will be viewed as being in the initial, or attached state, and will be said to be in state 0. 

We also assume that every minimum of the energy is uniquely characterized by the corresponding combination of pentamers that are in state 1, i.e., by its combination of components that equal 1. Formally,
\begin{itemize}
\item[$A_3$]
if two minima of the energy $\hat\bfx$ and $\tilde\bfx$ are such that, for each $i$, either
$\hat x_i=\tilde x_i=1$ or $\hat x_i,\tilde x_i<1$, then $\hat\bfx=\tilde\bfx$.

\item[$A_4$]
if two minima of the energy $\hat\bfx$ and $\tilde\bfx$ do not  belong to the same icosahedral orbit, then $E(\hat\bfx)\ne E(\tilde\bfx)$: in other words, each minimum (modulo symmetry) is characterized by a unique energy level.

\end {itemize}

\subsection{Icosahedral minima}

In this section we partially characterize the changes of the energy landscape resulting from variations of the bond strength $\gamma$.  We show that there is a critical threshold value for $\gamma$, such that below it the closed capsid is unstable. As $\gamma$, influenced by the environmental conditions, decreases below this critical value, the system undergoes a saddle-node bifurcation to an icosahedrally symmetric minimum which represents the final state. However, the analysis of the eigenvalues of the Hessian of the energy shows that the transition occurs by the activation of an icosahedrally symmetric 'breathing' mode that corresponds to an icosahedral expansion of the capsid, a result inconsistent with the energy cascade hypothesis, but consistent with normal mode analysis of capsid destabilization \cite{tama}.

First notice that assumption $A_3$ implies that there is at most equilibrium in which all pentamers are in the initial state, such that $x_i\ne 1$ for all $i$. Using the result (\ref{fixing2}) below, it follows that this equilibrium has full icosahedral symmetry,  i.e., 
$ x_i= x_0$, for all $i=1,\dots,12$.

The gradient and the Hessian of the energy are
\begin{equation}
\frac{\partial E}{\partial x_i}(\bfx)= f'( x_i) + 2\gamma\sum_{j=1}^{12} A_{ij} g_1( x_i,  x_j), \quad i=1,\dots,12.
\label{gradenergy0}
\end{equation}
and
\begin{equation}
\begin{split}
&\frac{\partial^2 E}{\partial x_i^2}(\bfx)= f''( x_i) + 2\gamma\sum_{j=1}^{12} A_{ij} g_{11}( x_i,  x_j),
\\
&\frac{\partial^2 E}{\partial x_i\partial x_j}(\bfx)=  2 \gamma A_{ij} 
g_{12}( x_i,  x_j).
\end{split}
\label{hessianenergy0}
\end{equation}
Then $ x_0$ must be a solution of the equation
\begin{equation}
h( x)=
f'( x) + 10\gamma g_1(x, x)=0.
\label{system1}
\end{equation}
The above equation has always the solution $x=1$, i.e., the configuration in which all pentamers have switched to state 1. Further, by $A_2$, the function $g_1( x, x)$ is nonnegative and $g_1(0,0)=g_1( 1, 1)=0$. Since $f'( x)<0$ for $x<1$, then there exists a critical value $\gamma_c$ such that    for $\gamma>\gamma_c$ there are  two solutions $ x_s= x_s(\gamma)<1$ and $ x_u= x_u(\gamma)<1$ of (\ref{system1}), with the property that 
$h'( x_s)>0$ and 
$h'( x_u)<0$, while for $\gamma<\gamma_c$ there are no solutions. Hence, the system undergoes a saddle-node bifurcation at $\gamma_c$.  

The Hessian of the energy computed at an isotropic state   $ x_i= x$ for all $i$ is
\begin{equation*}
aI +bA,
\end{equation*}
with $a=a( x)= f''( x) + 10\gamma 
g_{11}( x,  x)$ and $b=b( x)=2 \gamma g_{12}( x,  x)$. Notice that, by our hypotheses on $g,f$, we have that $b( x)\le0$, and the Hessian is proportional to the adjacency matrix $A$ of the graph (modulo the addition of a multiple of the identity). Therefore, its eigenspaces are also associated with the irreps of the representation $\widetilde{\rho}$ of $\mathcal{H}_3$ as in \eqref{H3_rep}. The eigenvalues of the Hessian listed in increasing order are  
\begin{center}
\begin{tabular}{|l|c|c|}
\hline
Eigenvalue & Dimension & Irrep \\
\hline
$\mu_1=a+5 b$ & 1 & $\rho_1$ \\
$\mu_2=a+ b \sqrt{5}$ & 3 & $\rho_2$ \\
$\mu_3=a-b$ & 5 & $\rho_5$ \\ 
$\mu_4=a- b \sqrt{5}$ & 3 & $\rho_3$ \\
\hline
\end{tabular}
\end{center}
Hence,
$$
h'( x)=a( x)+5b( x)=\mu_1,
$$
is the minimum eigenvalue of the Hessian, so that for $\gamma>\gamma_c$ the first critical point $ x_s$ is a relative minimum of the energy, since the minimum eigenvalue $\mu_1( x_s)$ of the Hessian is positive. A similar argument shows that  $ x_u$ is a saddle point. Notice also that, at the bifurcation point, the smallest eigenvalue of the Hessian changes sign, and the associated eigenspace is the isotropic line: hence, destabilization at the critical value $\gamma_c$ occurs by a breathing expansive mode.

%%%%%%%%%%%%%%%%%%%%%%%%%%

\section{Conformational changes}

In many cases, the configurational changes of the capsid are triggered by variations of the chemical environment of the virion, which modifies the interactions between the proteins and, by consequence, the energy function. In our simplified model, such changes are reckoned by variations of the parameter $\gamma$. 

However, the  saddle-node bifurcation occurring at $\gamma_c$, in which the destabilization occurs via an icosahedrally symmetric expansion, does not account for the complexity of the interactions between the capsomers. In fact, the cooperative nature of the stability of the capsid suggests that the destabilization occurs as a cascade of elementary events: the switching of single capsomers requires less energy than the simultaneous change of state of all of them, and once one of them has switched, its neighbors are destabilized and a cascade of destabilization event is triggered, with decreasing energy barriers as the number of unswitched pentamers decreases. 

A possible way to study the destabilization cascade is to work in the stable regime, and to treat the process in terms of a path visiting the metastable states of the energy  according to the law of rare events, using Large Deviations Theory (LDT) for stochastic dynamical systems. In other terms, we assume that, near the initial-state equilibrium, additive fluctuactions of the system due to  changes of the environment are able to drive it out of equilibrium, with increasing probability as $\gamma\to\gamma_c$, and the actual conformational change  can be described by this stochastic process.

Hence, we assume from now on that $\gamma>\gamma_c$, and denote by $\bfx_0=( x_s,\dots, x_s)$ the icosahedral minimum of the energy $E$ for which $x_s<1$, and by $\bfx_1=( 1,\dots, 1)$ the final-state icosahedral minimum. In problems involving the opening of the capsid, $\bfx_0$ and $\bfx_1$ correspond to the closed and open capsid, respectively.

One way to study the stability of a complex structure is to construct a dynamical system whose attractors are the  local minima of the energy, i.e.,  to assume that deviations from equilibria satisfy a deterministic dissipative gradient-flow dynamics
\begin{equation}
\dot{\bfx} = -\nabla E(\bfx),
\label{ODE00}
\end{equation}
where the superposed dot denotes differentiation with respect to time $t\in \rr^+$, and $\bfx=\bfx(t)$ is the motion in the configuration space $\rr^{12}$. Since $\gamma>\gamma_c$, the system (\ref{ODE00}) has an attractor at $\bfx=\bfx_0$ (i.e.,  $ x_i= x_0= x_s$ $\forall i$): the initial state is stable for the dynamics (\ref{ODE00}),  and every small perturbation of this state tends to vanish over finite time intervals.

Now, let $\bfW = (W^{(1)}(t), \ldots, W^{(12)}(t))$ be a $12$-dimensional Brownian motion defined on a probability space $(\Omega, \mathcal{A},\mathbb{P})$. We consider random perturbations of the dynamical system \eqref{ODE00}, by superimposing an additive noise  $ d W^{(i)}$ to each face $i$ of the polyhedron $\mcC$. This results in the stochastic differential equation
\begin{equation}
 d \bfx= -\nabla E(\bfx)  d t + \epsilon   d \bfW,
\label{SDE00}
\end{equation}
with $\epsilon>0$ a small parameter.
We remark that the Langevin dynamics (\ref{SDE00})  is here regarded only as a tool to explore the energy landscape of the system: the actual dynamics of the conformational change must involve more refined physical considerations. In the overdamped Langevin equation $\epsilon^2$ is proportional to the temperature.

\subsection{Results from large deviations theory}\label{sec:ldt}

In this section we briefly review some key concepts from LDT that we are going to use throughout the paper. We refer to \cite{freidlin} for the general theory. 

The main object of LDT is the action functional $S_T(\bfphi)$ which, for the equation \eqref{SDE00}, is given by
\begin{equation}\label{functional}
S_T(\bfphi) = \int_0^T |\dot\bfphi + \nabla E(\bfphi)|^2 dt,
\end{equation}   
where $T > 0$ and $\bfphi : [0, T] \rightarrow \R^{12}$ is an absolutely continuous path (actually, the functional is  defined on continuous paths, but its value is $+\infty$ if $\bfphi$ is not absolutely continuous). The LDT states that, given a bounded set $D \subseteq \R^{12}$, then the probability $\bbP^{\bfx}(\bfx(T) \in D)$ that the solution $\bfx(T)$ of \eqref{SDE00} belongs to $D$ given that $\bfx(0) = \bfx$  satisfies
\begin{equation*}
 \lim_{\epsilon \to 0} \epsilon^2 \ln \bbP^{\bfx}(\bfx(T) \in D) = -\min_{\bfphi \in \mcC_{\bfx,D}} S_T(\bfphi),
\end{equation*}
where $ \mcC_{\bfx,D} = \{ \bfphi \in \mcC([0,T], \R^{12}) : \bfphi(0) = \bfx, \bfphi(T) \in D \}$. If the event occurs, then $\bfx(t)$ is arbitrarily close to the minimizer 
\begin{equation}\label{minimizer}
\bfphi^* = \argmin_{\bfphi \in \mcC_{\bfx,D}} S_T(\bfphi),
\end{equation}
in the sense that, for every $\delta > 0$,
\begin{equation*}
\lim_{\epsilon \to 0} \bbP^{\bfx}\left(\sup_{0\leq t \leq T}|\bfphi^*(t)-\bfx(t)| < \delta \Big| \bfx(T) \in D \right) = 1.
\end{equation*}
Another central object in LDT is the quasipotential 
\begin{equation}\label{quasipot}
V(\bfx,\bfy) = \inf_{T>0} \min_{\bfphi \in \mcC_{\bfx,\bfy}} S_T(\bfphi),
\end{equation}
where $\mcC_{\bfx,\bfy} = \{ \bfphi \in \mcC([0,T],\R^{12}) : \bfphi(0) = \bfx, \bfphi(T) = \bfy \}$. For large time intervals, if $\bfx_0$ is the unique stable point of \eqref{ODE00}, the density $p(\bfx)$ associated with the stationary distribution of \eqref{SDE00}, provided it exists, is such that
\begin{equation}\label{stat_distr}
p(\bfx) \asymp \frac{1}{Z}\exp\left(-\frac{V(\bfx_0,\bfx)}{\epsilon^2}\right),
\end{equation}
where $Z$ is a normalization constant. Here $\asymp$ denotes the log-asymptotic equivalence for $\epsilon \to 0$, i.e. $\lim_{\epsilon \to 0} \epsilon^2\ln(p(\bfx)) = -V(\bfx_0, \bfx)$. 
For a gradient system like \eqref{SDE00}, the quasipotential at a point $\bfx$ lying in the basin of attraction of a minimum $\bfx_0$ of $E$ is given by 
\begin{equation}\label{quasipotgrad}
V(\bfx_0,\bfx) = 2(E(\bfx)-E(\bfx_0)).
\end{equation}

Suppose $\bfx_A$ and $\bfx_B$ are two (local) minima of $E$ separated by a single saddle point $\bfx_S$. Then the minimizer $\bfphi^*$ of the action functional $S_T(\bfphi)$ with $\bfphi(0) = \bfx_A$ and $\bfphi(T) = \bfx_B$ is the path 
such that, up to a normalization constant, 
\begin{equation}\label{mep}
(\nabla E)^\perp(\bfphi^*) = \bfzero,
\end{equation}
where $(\nabla E)^\perp$ is the component of $\nabla E$ normal to $\bfphi^*$:
\begin{equation}\label{eq:perp}
(\nabla E)^\perp(\bfphi^*) = \nabla E(\bfphi^*) - \langle \nabla E(\bfphi^*),\widehat{\bftau} \rangle \widehat{\bftau},
\end{equation}
where $\widehat{\bftau}$ is the unit tangent of $\bfphi^*$ and $\langle , \rangle$ denotes the standard Euclidean product. The path $\bfphi^*$ is referred to as the Minimum Energy Path (MEP) connecting $\bfx_A$ and $\bfx_B$ \cite{VDE1}. 

The dynamics of \eqref{SDE00} on the energy landscape can be modeled as a discrete time Markov chain with states  the minima of $E(\bfx)$ \cite{freidlin, cameron1}.  In this framework, two minima $\bfx_\alpha$ and $\bfx_\beta$ are connected 
if the  minimizer of $S_T(\bfphi)$ corresponds to the MEP $\bfphi^*$ with a single maximal value of the energy along it. The off-diagonal entries of the generator matrix $Q$ of the chain are given by
\begin{equation}\label{mat_gen}
Q_{\alpha\beta} = \left\{ \begin{array}{ll}
\displaystyle\exp\left(-\frac{2}{\epsilon^2}(E(\bfx_{\alpha\beta})-E(\bfx_\alpha))\right) & \text{if $\bfx_\alpha$ and $\bfx_\beta$ are connected} \\
\displaystyle 0 & \text{otherwise}
\end{array} \right.
\end{equation}
where $\bfx_{\alpha \beta}$ is the unique saddle point between $\bfx_\alpha$ and $\bfx_\beta$. The diagonal entries are chosen such that the sum of each row is zero:
\begin{equation*}
Q_{\alpha\alpha} = -\sum_{\beta \neq \alpha} Q_{\alpha\beta}.
\end{equation*}
The jump matrix $\Pi$  is obtained from $Q$ by setting
\begin{equation}\label{jump_matrix}
\Pi_{\alpha\alpha} = 0, \qquad \Pi_{\alpha\beta} = - \frac{Q_{\alpha\beta}}{Q_{\alpha\alpha}}, \quad \alpha \neq \beta.
\end{equation}
The jump matrix $\Pi$ is the generator matrix of a discrete Markov chain  \cite{prob}.  Following \cite{cameron1}, we introduce the limiting jump matrix (or zero-temperature jump matrix) 
\begin{equation}\label{lim_jump}
\Pi_0 = \lim_{\epsilon \to 0} \Pi.
\end{equation} 
The limiting jump matrix depends only on the values of the potential at the saddles \cite{cameron1}.

\subsection{Admissible states}

The energy landscape is complex and there are many local minima,  among which there is  the icosahedral minimum. By assumption $A_3$, local minima are completely characterized by the combination of pentamers $i$  such that $ x_i=1$, and $0< x_j<<1$ for the remaining indices, i.e., by the combination of detached pentamers. In this section we make precise this notion.

We parametrize the minima by strings $\bfs\in\{0,1\}^{12}$, where $s_i=1$ means that the pentagon $i$ is in state 1, for instance detached from the capsid, while $s_i=0$ means that the pentagon is still in its initial state, for instance attached to its neighbors.  Precisely, we say that a state $\bfs$ is \emph{admissible}
if it  belongs to the basin of attraction of a critical point $\hat\bfx(\bfs)$ of the  dynamical system (\ref{ODE00})  that corresponds to the same combination of  pentamers in state 1 as $\bfs$, i.e., if 

\begin{equation}
\bfs\in S\Leftrightarrow
\left\{
\begin{array}{l}
\displaystyle \lim_{t\to+\infty}\bfx(t, \bfs))=\hat\bfx(\bfs),
\\
s_i=1\quad
\Leftrightarrow\quad \hat x_i(\bfs)=1,
\\
s_i=0\quad
\Leftrightarrow\quad \hat x_i(\bfs)<1,
\end{array}
\right.
\label{admissibility}
\end{equation}
where $\bfx(t, \bfs)$ is the solution of (\ref{ODE00}) with initial conditions $\bfx(0)=\bfs$.

 We denote by $S$ the set of admissible states, and write $\sum_is_i$ for the number of pentamers in state 1. Special admissible states are the initial configuration $\bfs_0=(0,0,\dots,0)$  and the final configuration $\bfs_1=(1,1,\dots,1)$.

As an example, a set of admissible states for the energy (\ref{special_choice_energy}) is shown in Figure \ref{plott}. These  are obtained by  solving numerically the dynamical system (\ref{ODE00}) with initial conditions $\bfx(0)=\bfs$, to determine the attractor $\hat x(\bfs)$ whose basin $\bfx(0)$ belongs to.

As seen in Section \ref{sec:ldt}, the dynamics on the energy landscape can be analyzed by defining a Markov chain on the $S$ of admissible states. This analysis is carried out in the next section. 
 
\subsection{Transitions}\label{sec:trans}

In this section we introduce the basic tool that will allow to study transitions, i.e., the notion of connectivity between admissible states. The first basic idea here is that two states are connected if it is possible to go from one to the other by detaching pentamers. This automatically forbids reversible transitions in which pentamers can reattach or  switch back to their initial state. This hypothesis is consistent with the notion of 'locked' states proposed for the maturation pathway of HK97 \cite{ross}, but is also reasonable in the disgregation problem since detached capsomers cannot usually reattach to the capsid.  

The second basic notion is that two states are connected if the minimum energy path between them does not visit other minima, which is also a reasonable assumption here.

Precisely, given two admissible states $\bfs,\bfs'\in S$,  we say that $\bfs<\bfs'$  if $s_i'=1$ when $s_i=1$ and $\sum_is_i<\sum_is_i'$: in this case, since pentamers that are in state 1 in $\bfs$ are also in state 1 in $\bfs'$, the state $\bfs'$ is obtained by switching to $1$ some of the pentamers that are in state $0$ in $\bfs$.   The relation $<$ induces a strict partial order on the set $S$ consistent with the  structure of directed graph below.

We say that there is a directed edge between two admissible states $\bfs$, and $\bfs'$, and write $\bfs\to\bfs'$, if 
\begin{itemize}
\item[i)]
$\bfs' >\bfs$, so that no reverse transition or reattachment  of pentamers is allowed; and 

\item[ii)] there is a MEP connecting the corresponding minima $\hat\bfx(\bfs),\hat\bfx(\bfs')$, along which the energy has a single maximum, as described in Section \ref{sec:ldt}. 
\end{itemize}

We remark again that requirement (i) is strongly restrictive, in that it excludes conformational changes involving  pentamers going back to their initial configuration, or reattaching to the capsid. According to LDT, such transitions are indeed possible in the stochastic dynamics (\ref{SDE00}) and this may give rise to cycles. The Markov chain that we construct below, however, is meant to describe a restricted situation in which configurational changes are irreversible,  as is the case for most configurational changes in capsids, first of all the disgregation. In other terms, the values $x_i=1$ act as absorbing states for the $i$-th component of SDE (\ref{SDE00}).

 If $\bfx_m$ is the point of the MEP between $\hat\bfx(\bfs)$ and $\hat\bfx(\bfs')$ where the maximum is attained, we define the barrier between $\bfs$ and $\bfs'$ by
\begin{equation}
\beta(\bfs,\bfs')=V(\hat\bfx(\bfs),\bfx_m)=2(E(\bfx_m)-E(\hat\bfx(\bfs))),
\quad
\text{if}\quad \bfs\to\bfs',
\label{barrier}
\end{equation}
which is positive by construction. We set $\beta(\bfs,\bfs')=+\infty$ if $\bfs$ is not connected to $\bfs'$.

The above procedure endows the set $S$ with a structure of directed acyclic graph with positive weights.  To construct explicitly the weights $\beta$,  the MEP between any pair of states can be determined using the numerical procedure  introduced in \cite{VDE1, VDE2, VDE3},  taking for instance as initial path the straight line joining $\hat\bfx(\bfs)$ to $\hat\bfx(\bfs')$. 

\subsection{The Markov chain} 

In order to describe the dynamics on the energy landscape and determine the most probable transitions paths, as discussed in Section \ref{sec:ldt} we construct a Markov chain on the state space $S$. The generator matrix $Q$ of this chain is given by (compare with \eqref{mat_gen})
\begin{equation*}
Q(\bfs,\bfs')= \left\{
\begin{array}{ll}
\exp\left(-\frac{1}{\epsilon^2}\beta(\bfs,\bfs')\right)
& \text{if $\bfs\to\bfs'$},
\\
0& \text{otherwise},
\end{array}
\right.
\end{equation*}
and 
\begin{equation*}
Q(\bfs,\bfs)=-\sum_{\bfs':\bfs\to\bfs'}Q(\bfs,\bfs').
\end{equation*}
The jump matrix $\Pi$ as in \eqref{jump_matrix} is then given by
\begin{equation}
\Pi(\bfs,\bfs')= -\frac{Q(\bfs,\bfs')}{Q(\bfs,\bfs)}
=\frac{\exp\left(-\frac{\beta(\bfs,\bfs')}{\epsilon^2}\right)}{
\sum_{\bfs\to\bfs''}\exp\left(-\frac{\beta(\bfs,\bfs'')}{\epsilon^2}\right)
},
\qquad
\Pi(\bfs,\bfs)= 0.
\label{jump-matrix}
\end{equation}
Letting $\epsilon\to0$ we obtain the zero-temperature jump matrix (cf. \eqref{lim_jump})
\begin{equation}
\Pi_0(\bfs,\bfs')= \lim_{\epsilon\to0}\Pi(\bfs,\bfs').
\end{equation}
The matrix $\Pi_0$ defines a Markov chain on the set of admissible states $S$ by assuming that the transition probability between the states $\bfs,\bfs'$ is 
\begin{equation}
\pi(\bfs'|\bfs)=\Pi_0(\bfs,\bfs').
\end{equation}
The explicit form of the matrix $\Pi_0$ is the following:
\begin{equation}
\begin{split}
\Pi_0(\bfs,\bfs')&= \lim_{\epsilon\to0}\frac{1}{
\sum_{\bfs\to\bfs''}\exp\left(-\frac{(\beta(\bfs,\bfs'')-\beta(\bfs,\bfs'))}{\epsilon^2}\right)
}
\\
&=
\left\{
\begin{array}{ll}
\frac1{N(\bfs)}&\text{if }\bfs\to\bfs'\text{ and }\beta(\bfs,\bfs')=\min_{\bfs\to\bfs''}\beta(\bfs,\bfs'')
\\
0 &\text{otherwise}
\end{array}
\right.
\end{split}
\label{formulone1}
\end{equation}
where $N(\bfs)=|\{\bfs': \beta(\bfs,\bfs')=\min_{\bfs\to\bfs''}\beta(\bfs,\bfs'')\}|$   is the number of states that can be reached from $\bfs$ along a path with minimum barrier.  

The most probable transition path between the closed state $\bfs_0$ to the open state $\bfs_1$ is the realization of the chain that maximizes the transition probability at each step, i.e., the path along which the barriers are minimal among all the admissible transitions outgoing from each vertex (cf. Theorem 6.6.1 in \cite{freidlin}). 

We point out that the use of the zero-temperature matrix $\Pi_0$ in the construction of the Markov chain implies that we are only considering transitions between nearest neighboring states (which correspond to the first order cycles in the terminology of \cite{freidlin, cameron1}). The construction of higher order cycles would allow a wider analysis of the transitions between pentamers. However, the configuration space for the latter would become extremely complex, and the analysis carried out here would be unfeasible.    

\section{Reduction by symmetry}

In order to reduce the complexity of the problem, we will now restate the above results in terms of symmetry classes of minima. In fact, since the energy $E$ is invariant under the action of the symmetry group $\mathcal{H}_3$,  minima are mapped into minima by $\mathcal{H}_3$, and this induces a permutation action  on the set of minima of $E$. By (\ref{equivariant}) and uniqueness of the solution of (\ref{ODE00}), if $\bfx(t)$ is the solution of (\ref{ODE00}) with initial condition $\bfs$, then $H\bfx(t)$ is the solution of  (\ref{ODE00}) with initial condition $H\bfs$, for every $H\in\mathcal{H}_3$, and hence, by (\ref{admissibility}), it follows that
\begin{equation}
\bfs'=H\bfs\Leftrightarrow \hat \bfx(\bfs')= H\hat \bfx(\bfs),\qquad H\in \mathcal{H}_3,
\label{action_2}
\end{equation}
so that, in turn,  $\mathcal{H}_3$ acts on $S$.  Denoting by
\[
\text{Fix}(\bfs)=\left\{
H\in \mathcal{H}_3 : H\bfs=\bfs\right\}\]
the isotropy group of $\bfs$, (\ref{action_2}) implies that  $\hat\bfx(\bfs))$ is invariant under $\text{Fix}(\bfs)$, i.e., 
\begin{equation}
\hat\bfx(\bfs))=H\hat \bfx(\bfs),\qquad H\in \text{Fix}(\bfs),
\label{fixing}
\end{equation}
which in turn implies that $\hat\bfx(\bfs_0)$ and $\hat\bfx(\bfs_1)$ have all components equal, since $\text{Fix}(\bfs_0)=\text{Fix}(\bfs_1)=\mathcal{H}_3$ and therefore
\begin{equation}
\hat\bfx(\bfs_0)=H\hat \bfx(\bfs_0),
\quad
\hat\bfx(\bfs_1)=H\hat \bfx(\bfs_1),
\qquad \forall H\in \mathcal{H}_3,
\label{fixing2}
\end{equation}

The set $S$ therefore can be decomposed into orbits of $\mathcal{H}_3$. We denote by $\Sigma$ the set $S/\mathcal{H}_3$.  
Figure \ref{plott} shows the reduced state space $\Sigma$ for the energy (\ref{special_choice_energy}). We now show that the Markov chain $\Pi$ on $S$ induces a Markov chain $\tilde \Pi$ on $\Sigma$. 

Consider first a MEP $\bfphi^*$ connecting two minima $\bfx_0$ and $\bfx_1$: by construction, $\bfphi^*$ 
is a minimizer of the action functional $S_T(\bfphi)$ as in \eqref{functional}.
Since $E$ is invariant under $\mathcal{H}_3$, by (\ref{equivariant})  $\nabla E(H\bfx)=H\nabla E(\bfx)$ for $H\in \mathcal{H}_3$, and therefore, since $H$ is orthogonal, $S_T(\bfphi)=S_T(H\bfphi)$. This means that if $\bfphi^*$ is a MEP from $\bfx_0$ to $\bfx_1$, then $H\bfphi^*$ is a MEP from $H\bfx_0$ to $H\bfx_1$. Hence, noting that $\bfs<\bfs'$ implies $H\bfs<H\bfs'$,  
\begin{equation}
\bfs\to\bfs'\Rightarrow
H\bfs\to H\bfs'\qquad H\in \mathcal{H}_3.
\label{symmetry_connect}
\end{equation}
We now prove a basic property of barriers. From the invariance of $E$ and the above discussion, it follows that $E(\bfphi(s))=E(H\bfphi(s))$ for every $s\in[0,T]$, and using (\ref{action_2}) and (\ref{barrier}) this shows that 
\begin{equation}
\beta(H\bfs,H\bfs')=\beta(\bfs,\bfs'), \qquad H\in \mathcal{H}_3.
\label{fundamental_barrier}
\end{equation}
Notice in particular that, denoting by $\text{Fix}(\bfs)=\left\{
H\in \mathcal{H}_3 : H\bfs=\bfs\right\}$ the isotropy group of $\bfs$, 
\begin{equation}
\bfs\to\bfs'\Rightarrow
\bfs\to H\bfs'\quad\text{and}\quad
\beta(\bfs,H\bfs')=\beta(\bfs,\bfs'), \qquad H\in \text{Fix}(\bfs).
\label{fundamental_barrier_2}
\end{equation}
Hence, the value of the barrier between the state $\bfs$ and all states in $\text{Fix}(\bfs)\bfs'$ is the same. 

We  define a weighted graph with vertex set $\Sigma$ as follows: for $\bfsigma,\bfsigma'\in\Sigma$, we write
\begin{equation}
\bfsigma\to\bfsigma'
\quad\text{with weight }\quad \tilde\beta(\bfsigma,\bfsigma')
\end{equation}
if there exist $\bfs\in\bfsigma$, $\bfs'\in\bfsigma'$ such that
\begin{equation}
\bfs\to\bfs'
\end{equation}
in which case we write
\begin{equation}
\tilde\beta(\bfsigma,\bfsigma')= \min_{\bfs'\in\bfsigma'}\beta(\bfs,\bfs').
\end{equation}
The above definition is well given. In fact, let $\bfs'\in\bfsigma'$  such that  $\beta(\bfs,\bfs')=\min_{\bfs''\in\bfsigma'}\beta(\bfs,\bfs'')$, and 
assume that there exist $\bft\in\bfsigma$, $\bft'\in\bfsigma'$  such that  $\beta(\bft,\bft')=\min_{\bft''\in\bfsigma'}\beta(\bft,\bft'')\ne  \beta(\bfs,\bfs')$. Then by transitivity on the orbits there exists $H\in \mathcal{H}_3$ such that $\bft=H\bfs$, so that by (\ref{fundamental_barrier})
$\beta(\bft,\bft'')=\beta(\bfs,H^\top\bft')$.  Clearly, $H^\top\bft'\in \bfsigma'$ and this implies that $\beta(\bfs,\bfs')\le\beta(\bft,\bft'')$. Since the converse also holds, we arrive at a contradiction and $\beta(\bfs,\bfs')=\beta(\bft,\bft')$.

The above construction defines a Markov chain $\tilde\Pi$ on $\Sigma$:\begin{equation}
\tilde\Pi(\bfsigma,\bfsigma')
=\frac{\exp\left(-\frac{\tilde\beta(\bfsigma,\bfsigma')}{\epsilon^2}\right)}{
\sum_{\bfsigma\to\bfsigma''}\exp\left(-\frac{\tilde\beta(\bfsigma,\bfsigma'')}{\epsilon^2}\right)
},
\qquad
\tilde\Pi(\bfsigma,\bfsigma)= 0.
\label{jump-matrix2}
\end{equation}
To compute the low-temperature limit of (\ref{jump-matrix2}),  we make a further assumption on the system:

\begin{itemize}
\item[$A_5$] For every admissible state $\bfs$, if $\beta(\bfs,\bfs')=\beta(\bfs,\bfs'')$,  there exists $H\in\mathcal{H}_3$ such that $\bfs''=H\bfs'$.
\end{itemize}
This hypothesis is consistent with assumption $A_4$ on the minima of the energy, and is necessary to further reduce  the complexity of the problem, since it guarantees that, starting from a given state, there is a unique orbit to which the system can transform along a MEP.

Now,  as $\epsilon\to0$,  the dominant terms at the denominator of (\ref{jump-matrix2})  correspond to those $\bfsigma''$ such that $\tilde\beta(\bfsigma,\bfsigma'')=\min_{\bfsigma\to\bfsigma'''}\tilde\beta(\bfsigma,\bfsigma''')$. By assumption $A_5$, there can be only one orbit that realizes the minimum, so that the leading term only contains one summand. Hence, the low-temperature limit of (\ref{jump-matrix2}) is 
\color{black}
\begin{equation} 
%\begin{split}
\tilde\Pi_0(\bfsigma,\bfsigma')= \lim_{\epsilon\to0}\tilde\Pi(\bfsigma,\bfsigma')=
\left\{
\begin{array}{ll}
1&\text{if }\bfsigma\to\bfsigma'\text{ and }\tilde\beta(\bfsigma,\bfsigma')=\min_{\bfsigma''}\tilde\beta(\bfsigma,\bfsigma'')
\\
0 &\text{otherwise}
\end{array}
\right.
%\end{split}
\end{equation}

\section{Analysis of a special model}

\begin{figure}[h]
\centering
\includegraphics[scale=0.8]{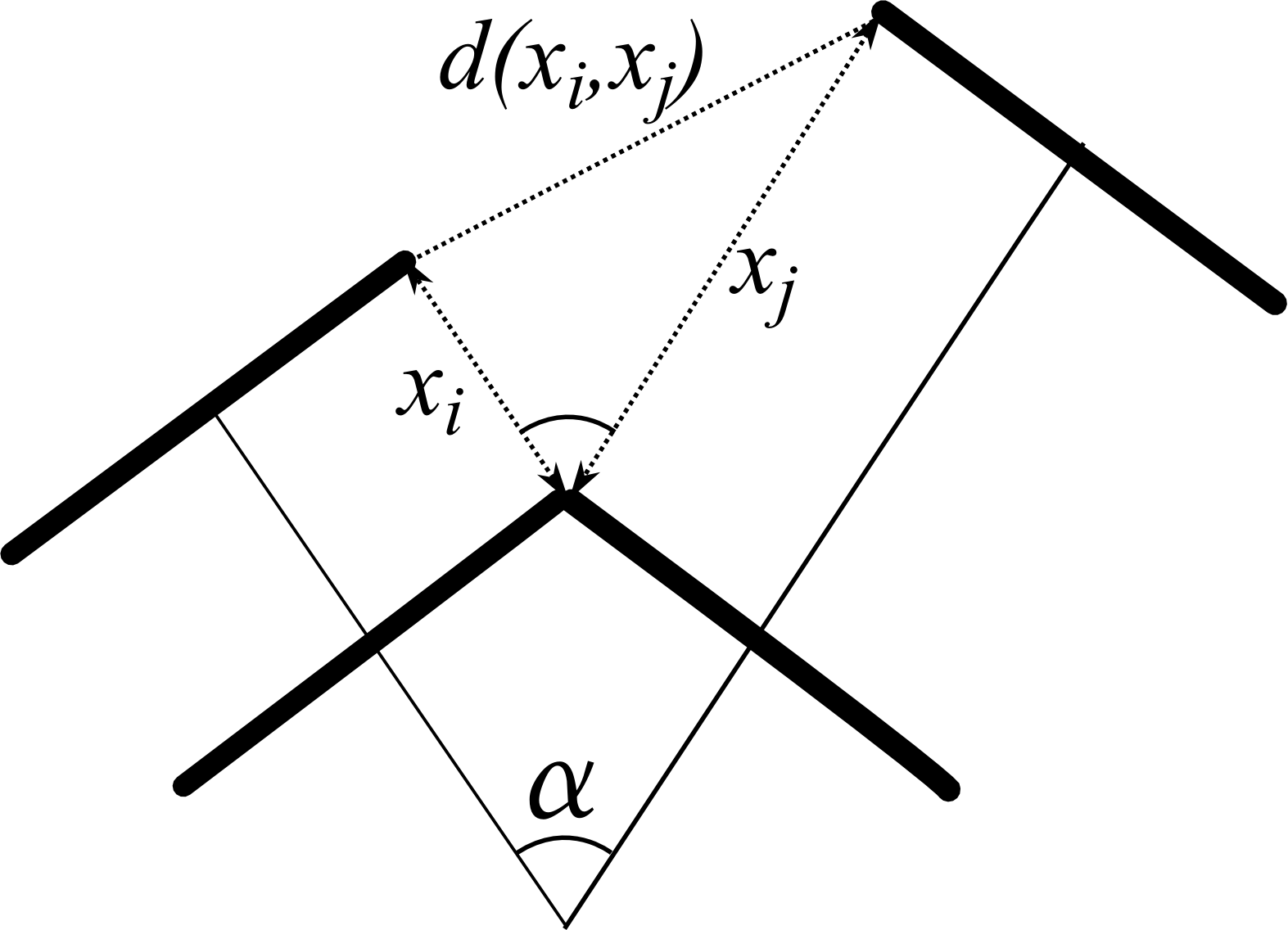}
\caption{The squared distance between two capsomers in the problem of the disgregation of the capsid. Schematic side view of two adjacent capsomers translating along their axis by $x_i$ and $x_j$.}
\label{cosalpha}
\end{figure}

We present below the analysis of a specific model, meant to describe the opening and disgregation of the capsid from the initial closed state and a final state in which all pentamers have detached from the capsid, motivated by the putative mechanism by which ERAV releases its genome within the host cell \cite{bakker}.  In this context, $x_i$ is the radial displacement of the capsomers along their axes (cf. Fig. \ref{cosalpha}). The model is 
based on the special form of energy function
\begin{equation}
E(\bfx) = \sum_{i=1}^{12} f(x_i) + \gamma\sum_{i,j=1}^{12} A_{ij} g(d(x_i, x_j)), \quad \gamma \in  \rr,
\label{energy01}
\end{equation}
where $f$ is the expansion energy and $g$ is the attractive interaction energy, that depends on the squared distance $d : \R^2 \rightarrow \R$ between the capsomers, defined by
\begin{equation}
d(x, y) =x^2 + y^2-2xy \cos\alpha,
\label{distance}
\end{equation}
where $\alpha$ is the angle between two neighboring icosahedral axes (cf. Figure \ref{cosalpha}).  

Precisely, we choose the expansion and attractive energies, and the constant $\gamma$ as
\begin{equation}
f(x)=\left\{
\begin{array}{ll}
(1-x)^2 & x\le1
\\
0 &x>1
\end{array}
\right.,
\qquad
g(x)=-e^{-ax},
\qquad
\gamma=0.1,\quad a=100.
\label{special_choice_energy}
\end{equation}
The expansion energy $f$ vanishes for $x\ge1$, while the attractive energy $g$ has a much shorter radius.   The attractive energy with fast exponential increase is consistent with the form proposed in \cite{kegel2004} (formula 1) for hydrophobic interactions between apolar surfaces,  while the expansive energy is just the simplest  choice of a polynomial monotone decreasing convex function consistent with hypothesis $A_1$.

\begin{figure}
        \centering
	\includegraphics[trim = {2.55cm 6cm 1.5cm 3cm}, clip, scale = 1]{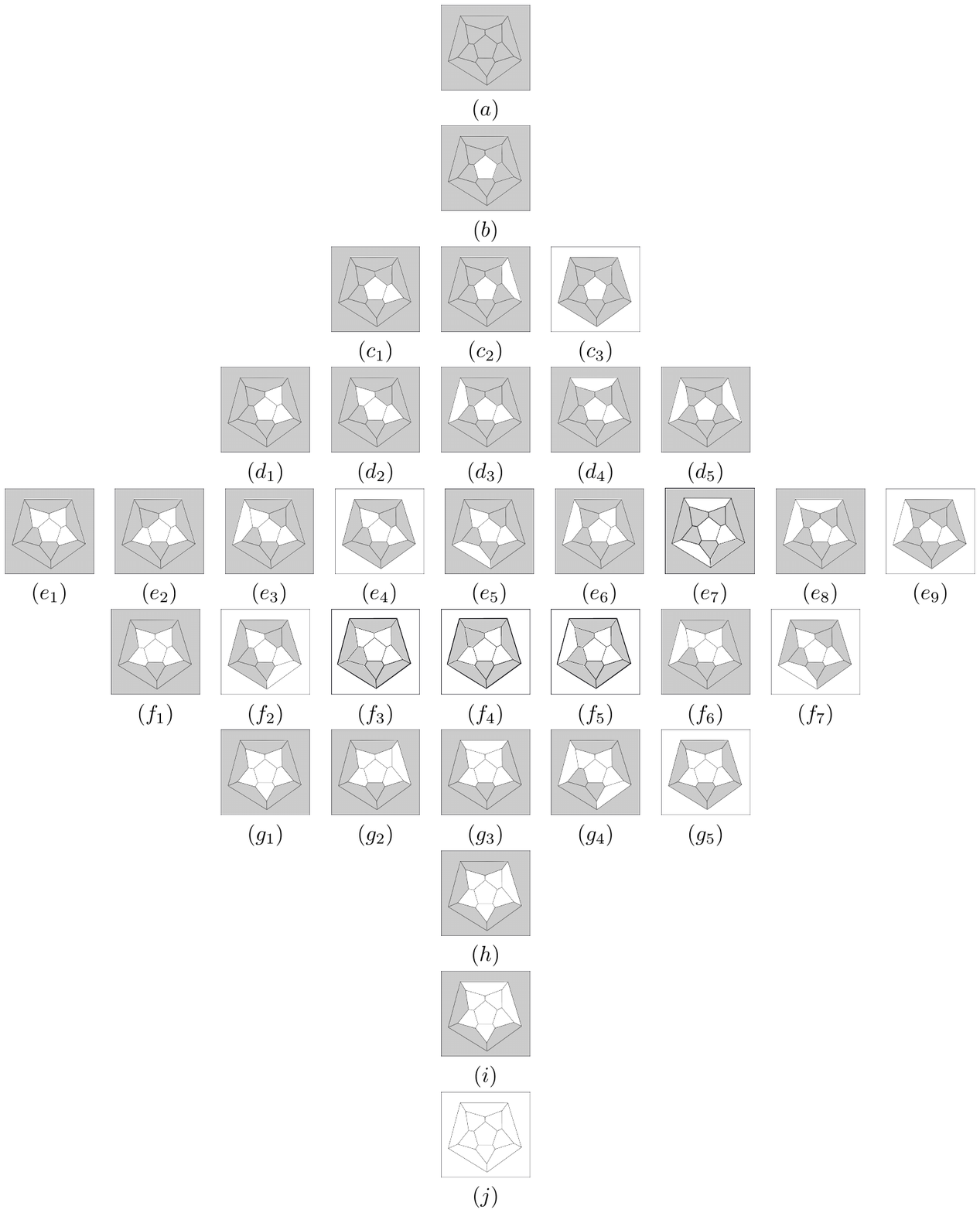}
	\caption{Representatives of the symmetry classes of admissible states  for the energy (\ref{special_choice_energy}). White pentagons represent detached pentamers. (a) is the closed capsid, (j) is the totally disgregated capsid.}
	\label{plott}

	\end{figure}

For each given number of detached pentamer, a representative of each icosahedral orbit of admissible states is shown in Fig. \ref{plott}.  Inspection of that figure  shows that

\begin{itemize}

\item There are multiple states with decreasing number of attached pentagons, but there is only one state (modulo the icosahedral group) with 5 and 4 attached pentamers. 

\item There are no admissible states with less than 4 attached pentamers. While it is obvious that a single pentamer cannot be attached to anything, even clusters of 4 or more pentamers cannot be stable in real capsids. This is a consequence of working with such a simplified model.                                                                         

\item In each admissible state the pentamer configuration is connected.

\end{itemize}

In Tables \ref{barriere1} and \ref{barriere2} below are listed the half-barriers $\beta(\bfsigma,\bfsigma')/2$ between symmetry classes of admissible states for the detachment problem and energy (\ref{special_choice_energy}). Only the half-barriers between nearest-neighbor states, that correspond to the detachment of a single pentamer, are indicated, with the exception  of some barriers between states that differ by at least two pentamers. These are written only when they are lower than the barriers between nearest neighbors, or when there is no stable nearest neighbor.

\begin{table}[h]

\begin{tabular}{ c | c | ccc|ccccc | ccccccccc | }
&$b$&$c_1$
&$c_2$&$c_3$&$d_1$&$d_2$&$d_3$&$d_4$&$d_5$&$e_1$&$e_2$&$e_3$&$e_4$&$e_5$&$e_6$&$e_7$&$e_8$&$e_9$
\\
\hline
$a$
&0.61&&&&&&&&&&&&&&&&&
\\
\hline
$b$
&&0.42&0.60&0.61&&&&&&&&&&&&&&
\\
\hline
$c_1$
&&&&&0.25&0.42&0.60&0.59&&&&&&&&&&

\\
$c_2$
&&&&& &0.24&0.42&0.41&0.59&&&&&&&&&
\\
$c_3$
&&&&& & &0.41& &&&&&&&&&&
\\
\hline
$d_1$
&&&&&&&&&&0.25&0.41&&&&0.59&&&
\\
$d_2$
&&&&&&&&&&0.09&0.24&0.41 &0.59&0.57&&&&
\\
$d_3$
&&&&&&&&&&&&0.23 &0.40&&0.24&&&0.41
\\
$d_4$
&&&&&&&&&&&0.08&&&0.40&0.25&0.57&0.41&

\\
$d_5$
&&&&&&&&&&& &&0.23&0.22& & &0.40&
\\
\hline
\end{tabular}

\caption{Half-barriers between states $a$ to $e_9$.}
\label{barriere1}
\end{table}

%%%%%%%%%%%%%%%%%%%%%%%%%%%%%%%%
%%%%%%%%%%%%%%%%%%%%%%%%%%%%%%%%
%%%%%%%%%%%%%%%%%%%%%%%%%%%%%%%%
%%%%%%%%%%%%%%%%%%%%%%%%%%%%%%%%
%%%%%%%%%%%%%%%%%%%%%%%%%%%%%%%%
%%%%%%%%%%%%%%%%%%%%%%%%%%%%%%%%

\begin{table}

\begin{tabular}{c | ccccccc | ccccc| c| c |c |}
&$f_1$&$f_2$&$f_3$&$f_4$&$f_5$&$f_6$&$f_7$&$g_1$&$g_2$&$g_3$&$g_4$&$g_5$&$h$&$i$&$j$
\\
\hline
$e_1$
&0.24&&0.57&&&0.41&&&&&&&&&
\\

$e_2$
&0.09&&&0.56&&0.24&&&&&&&&&
\\
$e_3$
&&&0.39&&&0.08&&&&&&&&&
\\
$e_4$
&&0.22&0.07&0.22&&&0.36&&&&&&&&
\\

$e_5$
&&&0.09&&&&0.37&&& 0.0484&&&&&
\\
$e_6$
&&&0.23&0.39&0.38&0.08&&&&&&&&&
\\
$e_7$
&&&&0.07&&&0.20&&&&&&&&
\\
$e_8$
&&&&&0.22&&0.37&&
  0.0475&&&&&&
\\
$e_9$
&&0.21&&&0.20&&&&&&&&&&
\\
\hline
$f_1$
&&&&&&&
&0.09&0.24&0.23&&0.55&&&
\\
$f_2$
&&&&&&&
&&&&0.07&&0.02&&
\\
$f_3$
&&&&&&&
&&&&&0.21&0.05&&
\\
$f_4$
&&&&&&&
&&&&0.07&0.06&&&
\\
$f_5$
&&&&&&&
&&&&&&&0.01&
\\
$f_6$
&&&&&&&
&&0.08&&0.38&&&&
\\
$f_7$
&&&&&&&
&&&&&&&&0.002
\\
\hline
$g_1$
&&&&&&&&&&&&
&0.23&&
\\
$g_2$
&&&&&&&&&&&&
&0.08&&
\\
$g_3$
&&&&&&&&&&&&
&0.08&&
\\
$g_4$
&&&&&&&&&&&&
&&&0.02
\\
$g_5$
&&&&&&&&&&&&
&&&0.02
\\
\hline
$h$
&&&&&&&&&&&&
&&0.07&
\\
\hline
$i$
&&&&&&&&&&&&
&&&0.05
\\

\hline
\end{tabular}
\caption{Half-barriers between states $f_1$ to $j$.}
\label{barriere2}

\end{table}

\begin{figure}[h]
\centering

\includegraphics[scale=0.6]{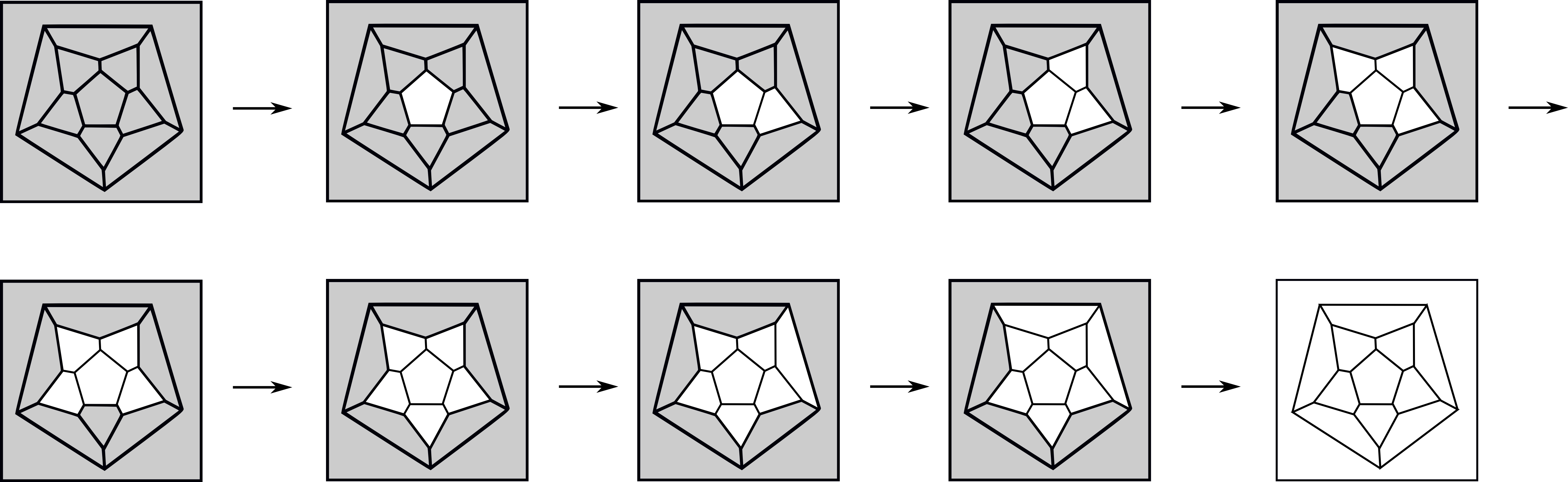}
\caption{Minimum energy path for the energy (\ref{special_choice_energy}). White pentagons are detached.}
\label{plott1}
\end{figure}

Inspection of Tables \ref{barriere1} and \ref{barriere2} shows  that the lowest barrier between a state and its out neighbors mostly corresponds to the detachment of one of the pentamers with the lowest number of attached neighbors.  Computing the non-zero entries of the zero-temperature shows that the most probable transition path from the closed to the open state is
\begin{equation}
a\to b  \to c_1 \to d_1 \to e_1 \to f_1 \to g_1 \to h\to i,
\end{equation}
withe the same labels as in Figure \ref{plott}.
 Figure \ref{plott1} shows a realization of the most probable transition path.

\section{Fluctuactions near equilibrium}

Linearization of the dynamical system (\ref{SDE00}) near equilibria gives information on the correlation between the fluctuactions of the building blocks of the system. 

Consider an admissible state $\bfs$ and the corresponding equilibrium $\hat\bfx(\bfs)$.  The Hessian 
$$K=K(\bfs)=\nabla\nabla E(\hat\bfx(\bfs)),
$$ 
of the energy at this point is positive definite by assumption (cf. \eqref{admissibility}), and, using (\ref{energy_inv}), (\ref{equivariant}) and (\ref{action_2}), we see also that 
\begin{equation}\label{equivariant2}
R^\top K(\bfs) R=K(\bfs),\qquad \forall R\in Fix(\bfs).
\end{equation}
Fluctuations near the stable equilibrium $\hat\bfx(\bfs)$ are ruled by the linearized system
\begin{equation}\label{linear1}
 d \bfx = - K\bfx  d t  + \epsilon   d \bfW,\qquad K=K(\bfs).
\end{equation}

Assuming the initial condition to be constant or normally distributed, the explicit solution of \eqref{linear1} is a Gaussian process given by (see \cite{arnold})
\begin{equation}\label{sol}
\bfx(t) = e^{-t K} \left( \bfx_0 + \int_0^t e^{z K}  d \bfW(z) \right).
\end{equation}
The expectation $\bfm =\bfm(t):= \mathbb{E}[\bfx(t)]$ and the covariance matrix $P(t)$ of $\bfx(t)$, are solutions of the linear ODEs
\begin{equation}\label{eqmedia1}
\left\{
\begin{array}{l}
\dot{\bfm} = - K \bfm,
\\
\dot{P} = - KP - PK + \epsilon^2 I.
\end{array}
\right.
\end{equation}
The solution of \eqref{eqmedia1}$_1$ is immediate and it is given by
\begin{equation}\label{media}
\bfm(t) = e^{-tK}\bfm(0).
\end{equation}
We find the explicit solutions of \eqref{eqmedia1}$_2$ using some results from matrix algebra.  We associate with a matrix $B \in GL(n,\R)$ the (column) vector $v(B) \in \R^{n^2}$ of its entries:
\begin{equation*}
v(B):= (B_{11},B_{12}, \ldots, B_{1n},B_{21}, \ldots, B_{nn})^T.
\end{equation*}
The following property holds for every $B,C,D \in GL(n,\R)$ \cite{barnett}:
\begin{equation}\label{prop}
v(C B D^T) = (C \otimes D) v(B),
\end{equation}
where $\otimes$ denotes the Kronecker product of matrices. We transform \eqref{eqmedia1}$_2$ into a vector equation obtaining
\begin{align*}
\frac{ d v(P(t))}{ d t} & = - v(KP)-v(PK) + \epsilon^2v(I) \\
& = - (K \otimes I ) v(P) - (I \otimes K) v(P) + \epsilon^2 v(I) \\
& = -(K \otimes I+ I \otimes K) v(P) + \epsilon^2  v(I),
\end{align*}
which is a linear non homogenous first order ODE, whose solutions is given by, 
setting $\widetilde{K} = K \otimes I  + I  \otimes K$,
\begin{equation}\label{vsol1}
v(P(t)) = e^{-t\widetilde{K}} \left( v(P(0))+ \int_0^t e^{z\widetilde{K}}v(I )  d z \right).
\end{equation}
Since $K \otimes I $ and $I  \otimes K$ commute, we have, using the properties of the Kronecker product (see \cite{johnson}),
\begin{align*}
\text{exp}(\widetilde{K}) & = \text{exp}(K \otimes I  + I  \otimes K) = \text{exp}(K \otimes I ) \text{exp}( I  \otimes K) \\
& = (\text{exp}(K) \otimes I ) (I  \otimes \text{exp}(K)) = \text{exp}(K) \otimes \text{exp}(K).
\end{align*}
Using properties \eqref{prop}, equation \eqref{vsol1} becomes
\begin{align*}
v(P(t)) & = (e^{-tK} \otimes e^{-tK}) \left( v(P(0)) + \epsilon^2 \int_0^t e^{zK} \otimes e^{zK} v(I )  d z \right) \\
& =  (e^{-tK} \otimes e^{-tK}) \left( v(P(0)) + \epsilon^2 \int_0^t v(e^{zK} I  e^{zK} )  d z \right) \\
& =   (e^{-tK} \otimes e^{-tK}) v\left( P(0) + \epsilon^2 \int_0^t e^{zK} I  e^{zK}   d z \right) \\
& = v \left\{ e^{-tK} \left(  P(0) + \epsilon^2  \int_0^t e^{2zK}   d z \right) e^{-tK} \right\}.
\end{align*}
Hence, the solution of \eqref{eqmedia1}$_2$ is given by
%\begin{equation*}
%P(t) =  e^{-tK} P(0) e^{-tK}   + \epsilon^2 \int_0^t e^{2(z-t)K}  d z 
%\end{equation*}
%i.e.,
\begin{equation}\label{solcov2}
P(t) =  e^{-tK} P(0) e^{-tK}   + \frac12\epsilon^2 K^{-1}(I-
e^{-2tK}) .
\end{equation}
The solution $\bfx(t)$ of \eqref{linear1} has therefore distribution $\mathcal{N}\left(\bfm(t),P(t)\right)$. Since $K$ is positive definite, we have 
$$
\lim_{t\to+\infty} \bfm(t)= \mathbf{0}, \qquad 
\lim_{t\to+\infty}P(t)= \frac12\epsilon^2 K^{-1}.
$$
Hence, $\bfx (t)$ converges in distribution to a Gaussian random variable $\bfX$ with mean $\bfm = \bfzero$ and covariance matrix $Q = \frac{\epsilon^2}{2} K^{-1}$. 

We point out that this result agrees with the LDT framework. In fact, the quasipotential of \eqref{linear1} as in \eqref{quasipot} in the basin of attraction of $\bfx_0$ is given by $V(\bfx) = \bfx \cdot K\bfx$. It follows from  \eqref{stat_distr} that the stationary distribution $p(\bfx)$ of \eqref{linear1} is asymptotically given by
\begin{equation}\label{stat_distr_lin}
p(\bfx) \asymp \frac{1}{Z}\exp\left( -\frac{V(\bfx)}{\epsilon^2}\right) =  \frac{1}{Z}\exp\left( -\frac{\bfx \cdot K\bfx}{\epsilon^2}\right),  
\end{equation}
which is the density of a Gaussian distribution $ \mathcal{N}\left(\mathbf{0},\frac{1}{2} \epsilon^2 K^{-1}\right)$. Here $Z$ is a normalization constant. 

The limiting distribution $p(\bfx)$ has important consequences. Recall that, given three continous random variables $X$, $Y$ and $Z$ with joint density distibution $p_{XYZ}(x,y,z)$, we say that $X$ is \emph{conditionally independent of $Y$ given $Z$}, and write $X \perp Y | Z$, if and only if
\begin{equation*}
p_{XY|Z}(x,y | z) = p_{X|Z}(x|z)p_{Y|Z}(y|z).
\end{equation*}
Let  $\bfX = (X_1, \ldots, X_n)  \sim \mathcal{N}(\bfm, Q)$ be a multivariate normal distribution. The \emph{concentration matrix} $C$ of $\bfX$ is the inverse of the covariance matrix $Q$ (provided that $\det(Q) \neq 0$). The entries of $C$ measure the correlation between the components. In particular, the \emph{partial correlation coefficients} are given by  
\begin{equation}\label{correlation}
\rho_{ij|\mathcal{S} \setminus \{i,j\}} := - \frac{(Q^{-1})_{ij}}{\sqrt{(Q^{-1})_{ii}(Q)^{-1}_{jj}}}. 
\end{equation} 
where $\mathcal{S} = \{1, \ldots, n \}$. Moreover, the following holds:
\begin{equation}\label{cond_indip}
X_i \perp X_j | \bfX_{\mathcal{S}\setminus \{i,j\}} \iff (Q)^{-1}_{ij} = 0.
\end{equation} 
If condition \eqref{cond_indip} is satisfied, then $\bfX$ is known as a \emph{Gaussian Markov random field} \cite{graphmodels}. 
In our case, the concentration matrix of $\bfx$ is $\frac{2}{\epsilon^2} K$, and we have, from \eqref{correlation}
\begin{equation}
\rho_{ij|\mathcal{S} \setminus \{i,j\}} = - \frac{K_{ij}}{\sqrt{K_{ii}K_{jj}}}.
\label{temp3}
\end{equation}
 Now, from the expression (\ref{hessianenergy0}), it follows that 
\begin{equation}
 \text{if  $s_i=\hat x_i=1$ then }\quad  K_{ij}(\bfs)=0, \quad\text{for }j\ne i,
\label{Kij}
\end{equation}
since, by $A_2$,  $g(x,y)$ and all its derivatives vanish whenever one of its arguments is 1. 
Moreover, if $\hat x_i<1$, so that $s_i=0$, then
\begin{equation}
K_{ii}(\bfs)= f''(\hat x_i) + 2\gamma\sum_{j:s_j=0} A_{ij} 
g_{11}(\hat  x_i,\hat  x_j).
\label{Kii}
\end{equation}
A first consequence of  (\ref{hessianenergy0}), (\ref{temp3}) and (\ref{Kij}) is that, at an admissible state, two pentamers are correlated if and only if they are adjacent in the configuration $\bfs$. In other words, the random variable $\bfx$ is a Gaussian graphical model with graph the subgraph of the icosahedral graph  induced by the vertices with zero components of $\bfs$.

In turn,  (\ref{Kii}) shows that the diagonal elements $K_{ii}$ of the Hessian  
depend only on the pentamers adjacent to $i$ in the state $\bfs$. Actually, (\ref{Kii}) suggests that $K_{ii}$ is larger the greater is the connectivity of pentamer $i$ in the state $\bfs$.  Hence, a pentamer with high connectivity has a small correlation coefficient  with its neighbors.

\section{Conclusions}

Large deviations theory for stochastic differential equations is based on the notion that arbitrarily small stochastic perturbations are able to lead any system out of equilibrium, over sufficiently long times, and minimum energy paths allow to determine the most probable transition paths between metastable states: the resulting Markov chain on the set of minima completely describes the stochastic dynamics on the energy landscape.  In this work we have used this approach to get insights into the process by which viral capsids change configuration, in either maturation or infection.

Our analysis supports the conjecture that destabilization occurs by a cascade of local events \cite{cardone,ross}: in fact, the energy is the sum of a destabilizing term on each pentamers, which does not depend on its connectivity, and a cohesive term opposing the transition, which accounts for the interactions between adjacent pentamers. Destabilization is the result of the competition between these terms: the expansive energy decreases whenever a pentamer changes state, but this requires providing an amount of cohesive energy proportional to the number of bonds broken in the  process. This suggests that the cascade occurs by destabilization of those pentamer that have less bonds, and therefore have to pay less energy to detach.  Simulations  confirm the intuitive picture above: for instance, in the parameter range in which we are working, whenever a pentamer has only one bond left, it switches to its final  configuration, because the energy gained by this process is larger than the amount lost by bond breaking.

For simplicity, we have computed in this paper just the zero-temperature transitions, i.e.,  the limit of the transition matrix of the Markov chain as $\epsilon\to0$, but the procedure is fully general and allows one to study all transitions between metastable states.

In order to further explore how random fluctuactions affect the stability of a complex interacting structure, we have also focused on the linearization of the system near a metastable state. Clearly, the destabilization dynamics is strongly nonlinear, but the analysis of  its linearization in the neighborhood of an attractor yields interesting information. First of all, the process is Gaussian and its limit distribution is a multivariate normal.  The concentration matrix $C$ is proportional to the Hessian of the energy at equilibrium, which has nonzero entries only when two pentamers are adjacent.  This is a first confirmation of the effect of locality on this model, but the point is that this allows to compute the conditional correlation coefficients between fluctuactions at adjacent pentamers. This shows that if a pentamer $i$ has many unbroken bonds and $H_{ii}$ and $C_{ii}$ are large, the fluctuactions are highly concentrated at that pentamer, and the correlation between this and adjacent pentamers is small. 

On the other hand, when a pentamer $i$ has a small connectivity, the concentration coefficient $C_{ii}$ is small, and the pentamer has large negative correlation coefficients with its neighbors. This means that fluctuactions tend to amplify, which is a clue of the destabilization effect.

The procedure described in this paper can be automatized and generalized to any system made of pairwise interacting building blocks, once the interactions are encoded in a simple graph.

\begin{acknowledgments}
We thank R. Kohn for suggesting the use of LDT to study transitions, and E. Vanden-Eijnden for valuable comments and suggestions.  GI acknowledges support by the Italian GNFM (Progetto Giovani 2016) and GI  and PC acknowledge support by the  by the University of Torino (research project 'Modelli Aleatori')).
\end{acknowledgments}

\newpage

\end{document}